\documentclass[conference]{IEEEtran}
\IEEEoverridecommandlockouts

\usepackage{cite}
\usepackage{amsmath,amssymb,amsfonts}
\usepackage{algorithm}
\usepackage{algorithmic}
\usepackage{graphicx}
\usepackage{textcomp}
\usepackage{xcolor}
\usepackage{subcaption}

\usepackage{colortbl}
\usepackage{dhucs} 
\usepackage{booktabs} 
\usepackage{multirow}
\usepackage{comment}
\usepackage{lineno}
\usepackage{color} 
\usepackage{setspace} 
\usepackage{graphicx} 
\usepackage{makecell}
\usepackage{enumitem} 
\usepackage{relsize}
\usepackage{tabularx}
\usepackage{diagbox}
\usepackage{amssymb}
\usepackage{bm}
\usepackage{xcolor}
\usepackage{colortbl}
\usepackage{hhline}
\usepackage{xurl}
\usepackage[colorlinks=false, pdfborder={0 0 0}]{hyperref}

\usepackage[a4paper, total={184mm,247mm}]{geometry}
\def\BibTeX{{\rm B\kern-.05em{\sc i\kern-.025em b}\kern-.08em
    T\kern-.1667em\lower.7ex\hbox{E}\kern-.125emX}}

\usepackage{siunitx}
\newcommand{\smalldeq}[1]{%
  {\small$#1$}%
}

\newcommand\korean[1]{}

\newcommand\refFigure[1]{Fig.~\ref{#1}}
\newcommand\refTable[1]{Table~\ref{#1}}

\newcommand\refAlgo[1]{Algorithm~\ref{#1}}
\newcommand\refEqn[1]{(\ref{#1})}

\newcommand\blind[1]{XXXX}

\begin{document}

\title{\LARGE\relsize{+1} TT-Edge: A Hardware–Software Co-Design for Energy-Efficient Tensor-Train Decomposition on Edge AI
}

\IEEEaftertitletext{\vspace{-2ex}}
\author{
Hyunseok Kwak$^{ \dag}$, Kyeongwon Lee$^{\dag}$, Kyeongpil Min, Chaebin Jung, and Woojoo Lee$^{\ast}$\\
School of Intelligent Semiconductor Engineering, Chung-Ang University, Seoul, Korea \\

\thanks{
This paper has been accepted for publication at the Design, Automation and Test in Europe (\textit{DATE 2026}). 
This document represents the camera-ready version.

$^{\dag}$ Hyunseok Kwak and Kyeongwon Lee contributed equally to this work.

$^{*}$Woojoo Lee is the corresponding author (space@cau.ac.kr).}
}

\maketitle

\begin{abstract}
The growing demands of distributed learning on resource-constrained edge devices underscore the importance of efficient on-device model compression. Tensor-Train Decomposition (TTD) offers high compression ratios with minimal accuracy loss, yet repeated singular value decompositions (SVDs) and matrix multiplications can impose significant latency and energy costs on low-power processors. In this work, we present \emph{TT-Edge}, a hardware–software co-designed framework aimed at overcoming these challenges. 
By splitting SVD into two phases—bidiagonalization and diagonalization, TT-Edge offloads the most compute-intensive tasks to a specialized \emph{TTD-Engine}. This engine integrates tightly with an existing GEMM accelerator, thereby curtailing the frequent matrix–vector transfers that often undermine system performance and energy efficiency. Implemented on a RISC-V-based edge AI processor, {TT-Edge} achieves a 1.7$\times$ speedup compared to a GEMM-only baseline when compressing a ResNet-32 model via TTD, all while reducing overall energy usage by 40.2\%. Notably, these gains come with only a 4\% increase in total power and minimal hardware overhead—enabled by a lightweight design that reuses GEMM resources and employs a shared floating-point unit.
Our experimental results on both FPGA prototypes and post-synthesis power analysis at 45\,nm demonstrate that {TT-Edge} effectively addresses the latency/energy bottlenecks of TTD-based compression in edge environments.
\end{abstract}



\section{Introduction}
The rapid growth of AI applications on edge devices has generated a strong demand for both privacy-preserving and domain-specific model development. 
A key paradigm meeting these requirements is distributed learning, where models are trained locally on edge devices, and only the resulting parameters—rather than raw data—are periodically shared and aggregated to update a global model \cite{Li:CVPR22, Cho:ISLPED2024, Yuan:IOT24}. 
Approaches like Decentralized Learning and Federated Learning exemplify this trend, offering clear advantages in safeguarding data ownership while allowing models to adapt to specific domains.

However, as distributed learning gains traction in real-world scenarios, the frequent exchange of model parameters between edge–cloud and edge–edge nodes has significantly increased communication overhead \cite{Zhou:TPDS21, Qin:IWC21, Liu:TSIPN22}. 
This bottleneck has prompted extensive research into on-device model compression techniques aimed at reducing the volume of parameters to be transmitted \cite{Yu:Network21, Song:IOT23, Khan:IOT24}. 
Among these strategies, tensor decomposition (TD) has received special attention for its ability to reshape high-dimensional tensors into more compact, lower-dimensional representations \cite{Dai:TNNL23, Yin:CVPR21}. 
Unlike methods that yield sparse parameter matrices, TD-based compression preserves small but dense matrix multiplications, making it more computationally efficient across diverse hardware.

\begin{table}[b]
    \caption{Simulation results comparing the performance of different TD methods for ResNet-32 on CIFAR-10.}
    \vskip -4pt
    \centering
    \resizebox{1\columnwidth}{!}{ 
    \begin{tabular}{cccc}
    \toprule
    {Method}                                & {Accuracy~(\%)}    & {Comp. ratio}   & {Final \#params}\\ \midrule
    {Uncompressed}                          & 92.49              & 1.0$\times$     & 0.47M\\ \midrule
    {Tucker Decomposition~\cite{Yin:AAAI22}}& 92.18              & 2.8$\times$     & 0.16M\\ 
    {TRD~\cite{Li:Springer21}}              & 91.44              & 2.7$\times$     & 0.17M\\ 
    {TTD (Focus of this work)}              & 92.09              & 3.4$\times$     & 0.14M\\ \bottomrule
    \end{tabular}
    }

    \label{tab:compression_ratio}
\vskip -2pt
\end{table}

Notable TD approaches include Tucker Decomposition~\cite{Yin:AAAI22}, Tensor-Ring Decomposition (TRD)~\cite{Li:Springer21}, and Tensor-Train Decomposition (TTD)~\cite{Ren:MM23}. 
TTD has garnered particular interest for achieving high compression ratios with minimal accuracy loss \cite{Yin:MLSys21, Ren:arxiv22} and for its compatibility with hardware accelerators \cite{Deng:ISCA19, Gong:ISCA23, Lee:DATE24}. 
In our study, we evaluated several TD-based methods on a lightweight ResNet model (ResNet-32 \cite{He:CVPR16}) trained on the CIFAR-10 dataset. 
As summarized in Table~\ref{tab:compression_ratio}, TTD attained a 3.4$\times$ compression ratio relative to the uncompressed model, while preserving 92.09\% accuracy. 
These promising results motivated our focus on TTD to mitigate communication overhead in distributed learning.

\begin{figure}[t]
    \centering
    \includegraphics[width=0.48\textwidth]{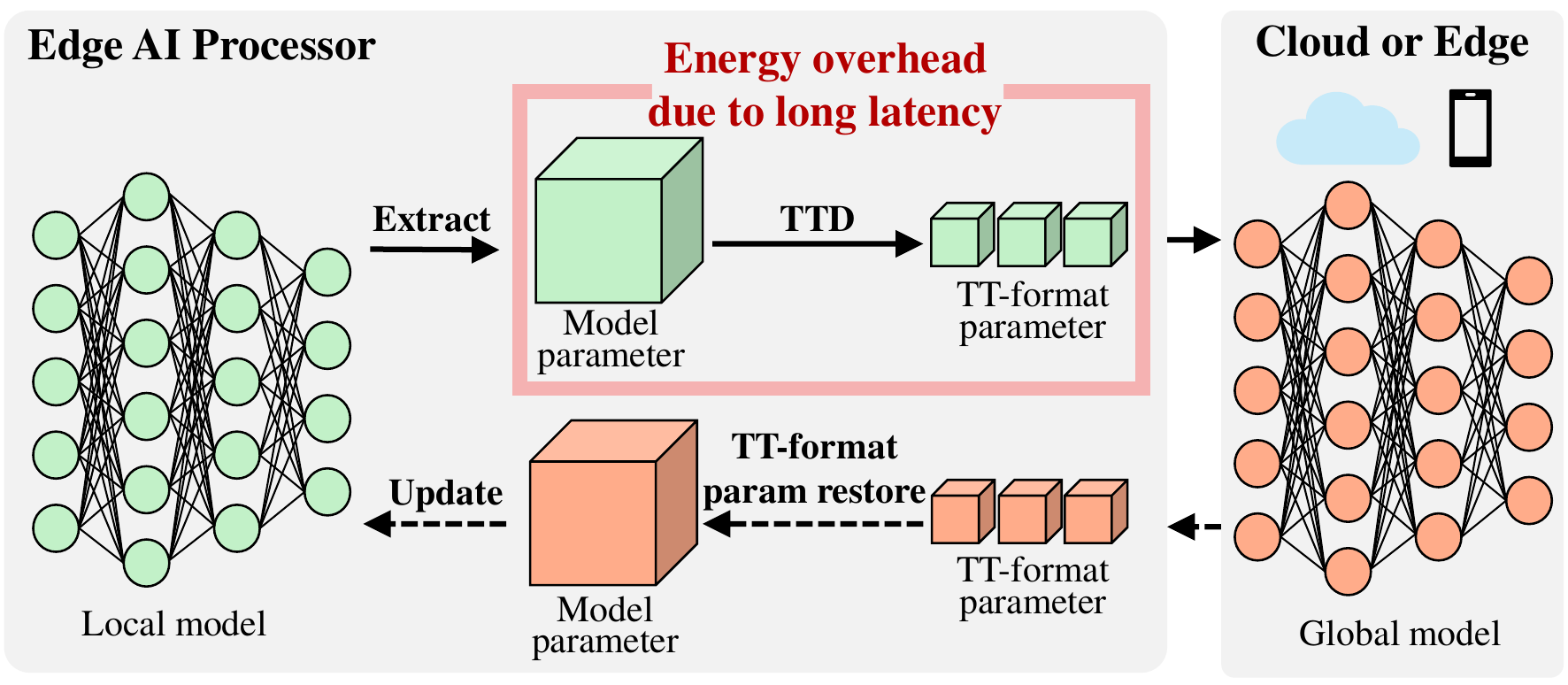}
    \vskip -2pt
    \caption{Compressed model parameter transmission via TTD.}
    \label{fig:FL_TTD}
   \vskip -4pt
\end{figure}

Despite its demonstrated effectiveness, deploying TTD-based compression on resource-constrained edge AI processors poses nontrivial challenges. 
\refFigure{fig:FL_TTD} illustrates a typical distributed learning workflow, in which each edge device compresses its local model parameters into TT-format before transmitting them for global model updates; the receiving node then reconstructs those parameters to update its local model. 
While the reconstruction phase of TTD has been optimized extensively for years—particularly in cloud-server training environments with abundant computational resources—the compression phase is comparatively understudied in the context of edge deployment.

On-device TTD compression presents two main hurdles. 
First, its repeated use of singular value decomposition (SVD) imposes high computation and memory demands—especially prohibitive for low-power edge AI processors \cite{Qu:TCAD21}. 
Second, repeated matrix multiplications during the compression process can overwhelm edge AI processors equipped with dedicated GEMM accelerators, creating high latency and energy overhead from frequent data transfers between the main core and the accelerator \cite{Qin:HPCA20}. 
Although hardware-based acceleration via QR iteration has been proposed \cite{Qu:TCAD21}, such solutions remain too resource-hungry for typical edge devices.

In this work, we introduce \textit{TT-Edge}, a framework that fundamentally rethinks TTD compression to better suit edge AI processors. 
Our approach consists of two key innovations:

\begin{enumerate}[leftmargin=*]
\item Efficient SVD through Bidiagonalization: Instead of relying on QR iteration, we adopt a bidiagonalization technique inspired by ScaLAPACK~\cite{Blackford:SIAM97}, breaking SVD into two phases (bidiagonalization and diagonalization). Profiling on edge AI processors revealed that the bidiagonalization phase dominates SVD computation—about 3.6$\times$ more time-consuming than diagonalization. We thus propose a specialized hardware accelerator for Householder-based bidiagonalization, carefully adapting the algorithm to edge constraints for seamless and efficient hardware integration.
\item TTD-Engine Architecture: To address the performance bottlenecks caused by frequent matrix–vector operations, we design a \textit{TTD-Engine} that integrates directly with existing GEMM accelerators. This allows matrix operations to continue benefiting from GEMM hardware while substantially reducing the latency and energy penalties typically incurred by shuttling data back and forth between the main core and accelerator.
\end{enumerate}

We implemented our TT-Edge framework in a RISC-V-based edge AI processor equipped with the proposed TTD-Engine. 
We designed the full RTL of the processor and performed FPGA prototyping to verify its functionality. 
Subsequently, we conducted performance evaluations on the prototype, confirming that the TT-Edge-enhanced processor achieved a 1.7$\times$ speedup compared to a conventional GEMM-based edge AI processor (i.e., baseline processor). 
Next, we synthesized our design using a 45nm PDK and carried out power simulations. 
Owing to TT-Edge’s lightweight design, the total processor power consumption increases by only 4\% relative to the baseline. 
Moreover, because the TTD-Engine offloads compression from the processor core—enabling partial clock gating when not actively needed—the power overhead during TTD runs is, on average, only about 1\% higher than a baseline in which the core remains continuously active.
Overall, our solution yields a 40.2\% energy reduction compared to the baseline, affirming that TT-Edge provides an efficient, low-power foundation for TTD-based model compression in distributed edge learning.


\section{TTD on Edge AI: Algorithm, Householder SVD Adaptation, and Key Challenges}
\subsection{TTD-based Model Compression}
\subsubsection{TTD Algorithm Overview}\label{subsec:ttd-compression}

TTD compresses an $N$-dimensional tensor 
\smalldeq{\bm{W} \in \mathbb{R}^{n_1 \times n_2 \times \dots \times n_N}}, composed of model parameters, 
into a set of 3D core tensors \smalldeq{\{\bm{G}_k\}_{k=1}^{N}}. Each core tensor 
\smalldeq{\bm{G}_k \in \mathbb{R}^{r_{k-1} \times n_k \times r_k}} satisfies boundary conditions \smalldeq{r_0 = r_N = 1}.
The general TTD algorithm is described in \refAlgo{alg:TTDecomp}; it iteratively performs:
\textit{Reshape}, \textit{SVD}, \textit{Sorting}, \textit{Truncation}, and a set of matrix multiplications. 
We outline the key steps below.

\paragraph{Reshape (line 7)} 
To begin, \smalldeq{\bm{W}} is reshaped into a matrix \smalldeq{\mathit{W}_{\text{temp}}}. 
For any tensor \smalldeq{\bm{X} \in \mathbb{R}^{I_1 \times \dots \times I_N}}, one can rearrange its dimensions 
to \smalldeq{[J_1, J_2, \dots, J_M]} provided that 
\smalldeq{\prod_{k=1}^{N} I_k = \prod_{m=1}^{M} J_m}. 
This operation preserves the overall ordering of elements while changing their dimensional layout.

\paragraph{SVD (line 8)}
Next, we compute the singular value decomposition (SVD) of \smalldeq{\mathit{W}_{\text{temp}}}, yielding 
\smalldeq{\mathit{W}_{\text{temp}} = U\,\Sigma\,V^{T}}. 
Here, \smalldeq{U \in \mathbb{R}^{M \times M}} and \smalldeq{V \in \mathbb{R}^{N \times N}} are orthogonal, whereas 
\smalldeq{\Sigma \in \mathbb{R}^{M \times N}} contains the singular values on its diagonal (for \smalldeq{M > N}).

\paragraph{Sorting (line 9)}
We then sort the singular values in \smalldeq{\Sigma} (via bubble sort in line 19 of \refAlgo{alg:TTDecomp}), forming \smalldeq{\Sigma_{s}}. In this process, the column indices of \smalldeq{\Sigma} are recorded in an index vector \textit{Ind}. Applying \textit{Ind} reorders \smalldeq{U} and \smalldeq{V^T} accordingly, yielding \smalldeq{U_{s}} and \smalldeq{V_{s}^{T}}.

\paragraph{Truncation (line 10)}
In line 29 (\textsc{\smalldeq{\delta}-Truncation}), we form the truncated matrices \smalldeq{U_{t}}, \smalldeq{\Sigma_{t}}, and \smalldeq{V_{t}^{T}}. 
Given a threshold \smalldeq{\delta}, let \smalldeq{i} be the smallest integer such that 
\smalldeq{\|\Sigma_{s}[\,i : \mathrm{rank}(\Sigma_{s}), :]\|_{F} < \delta,}
where \smalldeq{\|\cdot\|_{F}} denotes the Frobenius norm. Columns of \smalldeq{U_{s}} and rows of \smalldeq{V_{s}^{T}} beyond that index are discarded, retaining only the dominant singular values. Finally, we multiply \smalldeq{\Sigma_{t}} by \smalldeq{V_{t}^{T}} (lines 11--12) and reshape \smalldeq{U_{t}} into a new core tensor \smalldeq{\bm{G}_{k}}.


\paragraph{TTD Decoding}
Once the set of core tensors \smalldeq{\{\bm{G}_k\}_{k=1}^{N}} is obtained, 
the original tensor \smalldeq{\bm{W}} can be approximated via
\begingroup
\setlength{\abovedisplayskip}{2pt}   
\setlength{\belowdisplayskip}{0pt}   
\setlength{\abovedisplayshortskip}{0pt}
\setlength{\belowdisplayshortskip}{3pt}
\begin{equation}
\footnotesize
\bm{W}_{R} \;=\;
\bm{G}_1 \;\times_{1}\; \bm{G}_2 \;\times_{1}\; \dots \;\times_{1}\; \bm{G}_N,
\label{eq:TTDecoding}
\end{equation}
\endgroup
where \smalldeq{\times_{1}} denotes the tensor contraction operator.

In particular, for \smalldeq{\bm{X} \in \mathbb{R}^{I_1 \times \dots \times I_N}} 
and \smalldeq{\bm{Y} \in \mathbb{R}^{J_1 \times \dots \times J_M}} with \smalldeq{I_N = J_1}, 
let \smalldeq{\bm{T} = \bm{X} \times_{1} \bm{Y}}. 
Then, the contraction can be computed as:
\begingroup
\setlength{\abovedisplayskip}{2pt}
\setlength{\belowdisplayskip}{0pt}
\begin{equation}
\footnotesize
\begin{aligned}
\bm{T} &= \mathrm{Reshape}(\bm{X}, [\,I_1\,\dots\,I_{N-1}, I_N\,]) \cdot
          \mathrm{Reshape}(\bm{Y}, [\,J_1, J_2\,\dots\,J_M\,]), \\
\bm{T} &= \mathrm{Reshape}(\bm{T}, [\,I_1, \dots, I_{N-1}, J_2, \dots, J_M\,]),
\end{aligned}
\label{eq:Reshape}
\end{equation}
\endgroup
which computes matrix multiplication and reshapes the result back into a tensor~\cite{Oseledets:SIAM11}.
Repeating \refEqn{eq:Reshape} across all cores in \refEqn{eq:TTDecoding} 
reconstructs (approximately) the original parameters, completing the TTD-based compression and reconstruction procedure.


\begin{algorithm} [!t]
    \scriptsize
    \caption{Tensor Train Decomposition}\label{alg:TTDecomp}
    \begin{algorithmic}[1]        
    \STATE \textbf{function} TTD(\textbf{W}, $\epsilon$)
    \STATE \quad $\bm{W}$ : $Input\_Tensor \in \mathbb{R}^{n_1 \times \dots \times n_N}$
    \STATE \quad $\epsilon$ : prescribed accuracy
    \STATE \quad $\delta \leftarrow \frac{\epsilon}{\sqrt{d-1}}$$\times$$\lVert$\textbf{W}$\rVert _{F}$ : threshold of truncation  
    \STATE \quad $W_{temp} \leftarrow \bm{W}, r_0 \leftarrow 1$
    \STATE \quad \textbf{for} \textit{k} = 1 to \textit{N}-1 \textbf{do}
    \STATE \qquad $W_{temp} = Reshape(W_{temp}, [r_{k-1}n_{k}, \frac {numel(W_{temp})} {r_{k-1}n_{k}}])$
    \STATE \qquad \textit{U}, $\Sigma$, \textit{V\textsuperscript{T}} $\leftarrow$ SVD(\textit{W\textsubscript{temp}})
    \STATE \qquad $U_{s}, \Sigma_{s}, V_{s}^T \leftarrow$ Sorting\_Basis($U, \Sigma, V^T$)    
    \STATE \qquad $U_{t}, \Sigma_{t}, V^T_{t} \leftarrow$ $\delta$-Truncation($U_s, \Sigma_{s}, V_{s}^T, \delta$)
    \STATE \qquad $W_{temp} \leftarrow \Sigma_{t}V^T_{t} ,r_{k} \leftarrow rank(\Sigma_{t})$
    \STATE \qquad New Core: $\bm{G}_k \leftarrow Reshape(U_t, [r_{k-1}, n_k, r_k])$    
    \STATE \quad \textbf{end for}
    \STATE \quad $\bm{G}_N = Reshape(W_{temp}, [r_{N-1}, n_N, r_N])$ 
    \STATE \quad \textbf{return} $\bm{G}_1 ... \bm{G}_N$
    \STATE \textbf{end function}
    \STATE
    \STATE \textbf{function} Sorting\_Basis($U$, $\Sigma$, $V^T$)
    \STATE \quad $Ind$ : Bubble Sorted index array
    \STATE \quad $\Sigma_{s}, Ind \leftarrow Bubble\_Sort(\Sigma)$
    \STATE \quad \textbf{for} $i = 1$ to $rank(\Sigma)$ \textbf{do}
    \STATE \qquad $U_{s}[:,Ind[i]] \leftarrow  U[:,i], V^T_{s}[Ind[i], :] \leftarrow  V^T[i, :]$   
    \STATE \quad \textbf{end for}
    \STATE \quad \textbf{return} $U_{s}, \Sigma_{s}, V^T_{s}$    
    \STATE \textbf{end function}
    \STATE
    \STATE \textbf{function} $\delta$-Truncation($U_{s}$, $\Sigma_{s}$, $V^T_{s}$, $\delta$)
    \STATE \quad $k \leftarrow  \min\{i \in \{1,\dots,rank(\Sigma_{s})\} \mid \|\Sigma_{s}[i:rank(\Sigma_{s}),:]\|_{F}< \delta\}$
    \STATE \quad $U_{t}, \Sigma_{t}, V^T_{t} \leftarrow U_s[:, 1:k], \Sigma_{s}[1:k,1:k], V_{s}^T[1:k,:]$
    \STATE \quad \textbf{return} $U_{t}, \Sigma_{t}, V^T_{t}$
    \STATE \textbf{end function}
    \end{algorithmic}
\end{algorithm}

\subsubsection{Adapting Householder-based SVD Bidiagonalization}
\label{subsec:householder_svd}

SVD is a central step in TTD, as it decomposes a matrix into its singular values and corresponding basis matrices. To enable efficient SVD on resource-constrained platforms, we adopt a Householder Bidiagonalization (HBD) approach, which first reduces a matrix to an upper bidiagonal form and thereby lowers the complexity of the subsequent diagonalization~\cite{Lahabar:IPDPS09}.

\paragraph{Bidiagonalizing \smalldeq{\bm{A}}}
Given an \smalldeq{M \times N} matrix \smalldeq{A}, the HBD process produces an upper bidiagonal matrix \smalldeq{B}. 
For each \smalldeq{i = 1, 2, \dots, \min(M,N)}, we consider the submatrix 
\smalldeq{A[i\!:\!M,\; i\!:\!N]} and let \smalldeq{\bm{x} = A[i\!:\!M,\; i].} 
We form a Householder vec. \smalldeq{\bm{v}} by
\begingroup
\setlength{\abovedisplayskip}{2pt}
\setlength{\belowdisplayskip}{2pt}
\begin{equation}
\footnotesize
\begin{aligned}
\bm{v} 
  &= \bm{x} - \operatorname{sign}(\bm{x}_{1})\,\lVert \bm{x} \rVert \,\bm{e}_{1}, 
  \quad
  \bm{e}_{1} = \bigl[\,1,\,0,\dots,0\bigr]^T.
\end{aligned}
\label{eq:gen_v_left}
\end{equation}
\endgroup
Then, the Householder reflector is defined as
\begingroup
\setlength{\abovedisplayskip}{2pt}
\setlength{\belowdisplayskip}{2pt}
\begin{equation}
\footnotesize
H = I - 2\,\frac{\bm{v}\,\bm{v}^{T}}{\bm{v}^{T}\bm{v}},
\label{eq:HI1}
\end{equation}
\endgroup
which is applied from the left (\smalldeq{A \leftarrow H\,A}) to eliminate subdiagonal entries in \smalldeq{A[i\!:\!M,\; i\!:\!N]}. 
We refer to this step as the \emph{left Householder transform}.

Next, we focus on the first row of \smalldeq{A[i\!:\!M,\; i\!+\!1\!:\!N]}, denoted \smalldeq{\bm{y}}. 
We form another Householder vector:
\begingroup
\setlength{\abovedisplayskip}{2pt}
\setlength{\belowdisplayskip}{2pt}
\begin{equation}
\footnotesize
\begin{aligned}
\bm{v} 
  &= \bm{y} - \operatorname{sign}(\bm{y}_{1})\,\lVert \bm{y} \rVert \,\bm{e}_{1},
  \quad
  \bm{e}_{1} = \bigl[\,1,\,0,\dots,0\bigr],
\end{aligned}
\label{eq:gen_v_right}
\end{equation}
\endgroup
and compute
\begingroup
\setlength{\abovedisplayskip}{2pt}
\setlength{\belowdisplayskip}{2pt}
\begin{equation}
\footnotesize
H = I - 2\,\frac{\bm{v}^{T}\bm{v}}{\bm{v}\,\bm{v}^{T}}.
\label{eq:HI2}
\end{equation}
\endgroup
Applying \smalldeq{H} from the right (\smalldeq{A \leftarrow A\,H}) removes off-diagonal entries in that row. 
We call this the \emph{right Householder transform}. 
Repeating the above left and right transforms for \smalldeq{i = 1, \dots, N} yields an upper bidiagonal matrix \smalldeq{B}.

\paragraph{Forming \smalldeq{\bm{B}}}
By aggregating all left Householder reflectors, we obtain \smalldeq{U_{B}}, while those used on the right form \smalldeq{V_{B}^{T}}. Thus,
\smalldeq{A \;=\; U_{B}\,B\,V_{B}^{T}.}
\smalldeq{B} is the upper bidiagonal matrix.

\paragraph{Diagonalizing \smalldeq{\bm{B}}}
In the final step, \smalldeq{B} is diagonalized by a standard QR-based procedure:
\smalldeq{B \;=\; Q\,\Sigma\,Q^{T},}
where \smalldeq{Q} is orthonormal and \smalldeq{\Sigma} is a diagonal matrix of singular values. Combining these results, the SVD of \smalldeq{A} is expressed as
\smalldeq{A \;=\; U \,\Sigma\, V^{T},}
where \smalldeq{U = U_{B}\,Q} and \smalldeq{V^{T} = Q^{T} V_{B}^{T}}.

\subsection{TTD Compression Challenges in Edge AI Processors}
\label{subsec:problem_definition}

To identify the key challenges of performing TTD-based model compression on edge AI processors, we first assume a practical baseline architecture and then analyze the TTD compression workflow in detail. The baseline processor under consideration has the following main components:

\begin{itemize}[leftmargin=*]
    \item {GEMM Accelerator:} A hardware module supporting $16\times16$ matrix multiplication with a $320\,\text{KB}$ on-chip scratchpad memory (SPM).
    \item {Processor Subsystem:} A Rocket RISC-V core~\cite{Rocket} equipped with DDR3 DRAM, $128\,\text{KB}$ of system SRAM, and various peripherals.
    \item {DMA Engine:} An integrated DMA unit that enables high-throughput data transfers between the core, GEMM accelerator, and memory without constant core intervention.
\end{itemize}

On this baseline processor, TTD-based compression can be split into operations handled solely by the core and those where the core collaborates with the GEMM accelerator. For instance, HBD, one key step in TTD, consists of generating Householder vectors as defined in \eqref{eq:gen_v_left} and \eqref{eq:gen_v_right}, followed by creating and applying the corresponding reflection matrices in \eqref{eq:HI1} and \eqref{eq:HI2}. Taking the left transform as an example,
\smalldeq{
  2 \,\frac{\bm{v}}{\bm{v}^{T}\bm{v}}(\bm{v}^{T}A)
}
can be viewed as a combination of one vector–scalar division and two matrix multiplications. While Householder vector generation, scalar division, sorting, and truncation must be performed on the core (because the GEMM accelerator does not directly support them), large-scale matrix multiplications in TTD can be offloaded to the accelerator. However, the accelerator’s $16\times16$ processing limit requires dividing the entire matrix into multiple blocks—an approach known as blockwise matrix multiplication. In this scenario, the core is responsible for calculating each block’s address, dimensions, and data layout, sending these parameters to the accelerator, and transferring block data to the SPM.

These architectural constraints introduce several bottlenecks during TTD-based compression:

\begin{enumerate}[leftmargin=1.5em]
    \item \textbf{Limited operation support:} The GEMM accelerator does not handle Householder vector generation, scalar division, sorting, and truncation, leaving the core to perform all of these tasks and causing significant computational load and latency.
    \item \textbf{Communication overhead in blockwise multiplication:} Because the accelerator handles only $16\times16$ matrix blocks, the core frequently needs to compute block parameters and relay them to the accelerator, incurring cumulative latency from repeated communication.
    \item \textbf{Frequent data movement:} Repeatedly moving large blocks between DRAM and the SPM for every matrix multiplication step substantially increases memory and interconnect latency.
\end{enumerate}

In addition to these latency-related issues, such bottlenecks have a negative impact on power consumption and overall energy efficiency—key concerns in edge AI environments. The core must sustain a high operational load to manage non-GEMM operations, while also coordinating blockwise GEMM tasks and continuously interacting with the accelerator. Consequently, the system’s energy efficiency suffers due to the compounding overhead of intensive core activity and frequent accelerator communication.

In this paper, we present an integrated HW–SW approach for efficient TTD on resource-constrained edge AI processors.

\section{Proposed TT-Edge Solution}
\label{sec:proposed_solution}
To tackle the latency and energy challenges, we propose the \emph{TT-Edge} solution to mitigate the latency issues while also addressing power and energy consumption in edge AI processors. First, the main ideas for reducing latency can be summarized as follows:

\begin{enumerate}[leftmargin=1.5em]
    \item \textbf{Dedicated TTD compression accelerator:}  
    To reduce the core's high computational latency, we introduce a lightweight accelerator specialized for TTD-based compression. Offloading tasks such as Householder transformations and SVD operations from the core significantly lowers end-to-end latency.

    \item \textbf{Direct interconnection with the GEMM accelerator:}  
   To alleviate the repeated round trips through the system interconnect, we enable a direct connection between the TTD accelerator and the GEMM accelerator, allowing blockwise GEMM parameters to be computed on the new accelerator rather than the core. 
   
    \item \textbf{On-Chip retention of Householder vectors:}  
    A fundamental solution to memory-interconnect latency (e.g., redesigning the entire memory and network systems) lies beyond the scope of this TTD-focused study. Instead, we propose a practical approach: store the Householder vectors within the SPM during the bidiagonalization process, preventing repeated DRAM access for these vectors.
\end{enumerate}

Next, the ideas to address energy-related issues are:
\begin{enumerate}[leftmargin=1.5em]
    \item \textbf{Lightweight hardware design for energy efficiency:}  
    Beyond reducing latency, TT-Edge must also tackle power and energy constraints. Through a HW--SW co-design approach, we modify the TTD algorithm to reuse common hardware resources with minimal redundancy. This careful optimization keeps the accelerator’s power consumption low.

    \item \textbf{Maximizing core clock gating:}  
        To increase the time during which the core can remain clock-gated, the dedicated accelerator is designed to support as many TTD operations as possible. This approach maximizes offloading to the accelerator and further reduces power consumption.
    
    \item  \textbf{Integrating the existing GEMM accelerator into the dedicated accelerator:}  
    Since TTD and AI processes are mutually exclusive, the existing GEMM accelerator typically remains in standby during TTD execution. To take full advantage of its capabilities, we incorporate the GEMM accelerator inside our specialized design, enabling it to be fully utilized during TTD processing.
\end{enumerate}

\begin{figure}
  \vskip -10pt
    \centering
    \includegraphics[width=\columnwidth]{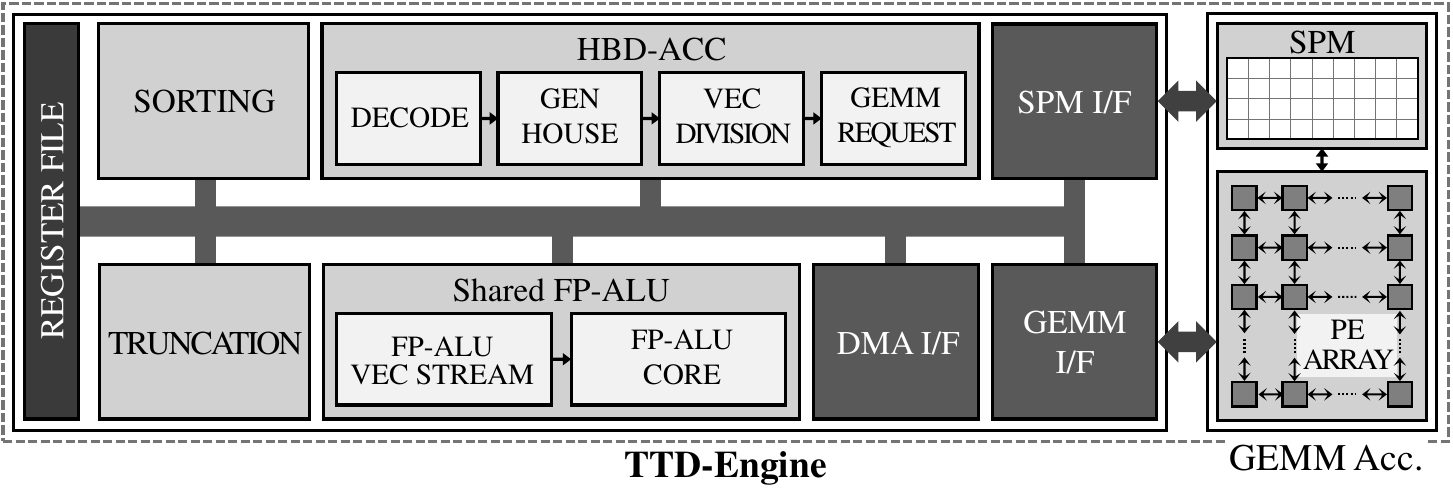}
    \vskip -6pt
    \caption{System-level architecture of the proposed TTD-Engine.}
    \label{fig:TTDE_arch}
\end{figure}

\begin{algorithm}[t]
    \scriptsize
    \caption{Proposed Householder Bidiagonalization for the HBD-ACC in the TTD-Engine.} \label{alg:adjusted_HBD}
    \begin{algorithmic}[1]
        \STATE \textbf{function} HOUSEHOLDER\_BIDIAGONALIZE(\textit{A})
        \STATE \quad Set $B$ to an $N \times N$ zero matrix
        \STATE \quad Set $U_{B}$, $V_{B}$ to $M \times N$, $N \times N$ identity matrix
        \STATE \quad \textbf{for} $i$ $=$ 1 to $N$ : \textbf{do}
        \STATE \qquad $B[i,i], v_{L}$ $\leftarrow$ HOUSE($A[i:M,i]$)
        \STATE \qquad HOUSE\_MM\_UPDATE($B[i,i], v_{L}, A[i:M,i+1:N], 0$)
        \STATE \qquad $A[i,i]$ $\leftarrow$ $v_{L}[1]$
        \STATE \qquad \textbf{if} $i < N$ \textbf{then}
        \STATE \qquad \quad $B[i,i+1], v_{R}$ $\leftarrow$ HOUSE($A[i,i+1:N]$)
        \STATE \qquad \quad HOUSE\_MM\_UPDATE($B[i,i+1], v_{R}, A[i+1:M,i+1:N], 1$)
        \STATE \qquad \quad $A[i,i+1]$ $\leftarrow$ $v_{R}[1]$
        \STATE \quad \quad \textbf{end if}
        \STATE \quad \textbf{end for}
        \STATE \quad \textbf{for} $i$ $=$ $N$ to 1 : \textbf{do}
        \STATE \qquad $v_{L}$ $\leftarrow$ $A[i:M,i]$, $v_{R}$ $\leftarrow$ $A[i,i+1:N]$
        \STATE \qquad HOUSE\_MM\_UPDATE($B[i,i], v_{L}, U_{B}[i:M,i+1:N], 0$)
        \STATE \qquad HOUSE\_MM\_UPDATE($B[i,i+1], v_{R}, V_{B}^{T}[i+1:N,i+1:N], 1$)
        \STATE \quad \textbf{end for}
        \STATE \quad \textbf{return} $U_{B}, B, V^{T}_{B}$
        \STATE \textbf{end function}
        \STATE
        \STATE \textbf{function} HOUSE($x$)
        \STATE \quad $v$ $\leftarrow$ $x$, \quad $q$ $\leftarrow$ $-sign(v[1]) * \lVert v \rVert$, \quad $v[1]$ $\leftarrow$ $v[1] + sign(v[1]) * \lVert v \rVert$
        \STATE \quad \textbf{return} $q, v$ 
        \STATE \textbf{end function}
        \STATE
        \STATE \textbf{procedure} HOUSE\_MM\_UPDATE($q, v, SubArray, order$)
        \STATE \quad $\beta$ $\leftarrow$ $v[1] * q$
        \STATE \quad $vec_{1}$ $\leftarrow$ (order = 0)? $v / \beta$ : $v \times SubArray^{T}$
        \STATE \quad $vec_{2}$ $\leftarrow$ (order = 0)? $v^{T} \times SubArray$ : $v / \beta$
        \STATE \quad $SubArray$ $\leftarrow$ $SubArray + vec_{1} \times vec_{2}$
        \STATE \textbf{end procedure}
    \end{algorithmic}
\end{algorithm}

\vspace{-1pt}
We collectively refer to this dedicated accelerator as the \emph{TTD-Engine}. 
\refFigure{fig:TTDE_arch} shows the overall TTD-Engine architecture, which is centered on the {HBD-ACC} module for accelerating Householder Bidiagonalization. Supporting components include the {Shared FP-ALU} for floating-point operations, as well as {SORTING} and {TRUNCATION} modules to complete the TTD workflow. In the following subsections, we detail the design of each module and demonstrate how they address the latency and energy bottlenecks.

\begin{figure}
    \centering
    \includegraphics[width=0.95\columnwidth]{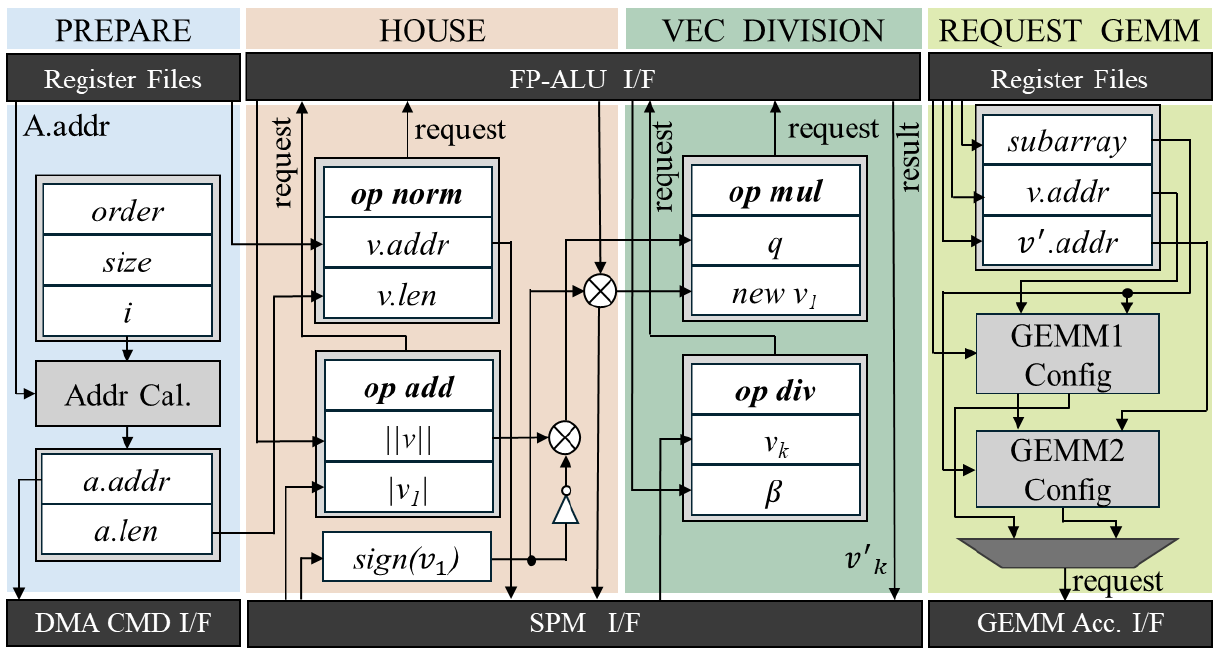}
    \vskip -4pt
    \caption{Architecture of the HBD-ACC.}
    \label{fig:HBD_Controller}
\end{figure}

\vspace{-4pt}
\subsection{Design of the HBD-ACC}
\label{subsec:hbd_acc}

The HBD-ACC is the core component for efficiently accelerating the HBD process with minimal resource usage. 
Because HBD involves different procedures (left or right transforms) depending on the input vector orientation, we first unify these steps into a single flow to enable reuse of low-power hardware blocks. \refAlgo{alg:adjusted_HBD} depicts our modified Householder transform procedure, combining:
\begin{itemize}[leftmargin=*]
    \item {Householder Reduction (lines 4--13):} Applies Householder transforms to the original matrix.
    \item {Householder Accumulation (lines 14--18):} Updates and accumulates the Householder matrices.
\end{itemize}
In this process, the \textit{HOUSE} function corresponds to generating Householder vectors (as in \refEqn{eq:gen_v_left} and \refEqn{eq:gen_v_right}), while the creation and application of reflection matrices are consolidated into a single procedure, \textit{HOUSE\_MM\_UPDATE}.

Leveraging this unified algorithm, we design the {HBD-ACC} to independently execute the entire HBD routine and minimize frequent interaction with the main core. \refFigure{fig:HBD_Controller} illustrates its architecture, which implements the \textit{HOUSE} and \textit{HOUSE\_MM\_UPDATE} steps in four stages: \texttt{PREPARE}, \texttt{HOUSE}, \texttt{VEC DIVISION}, and \texttt{REQUEST GEMM}.

During \texttt{PREPARE}, the address calculator determines the address for the Householder vector (\textit{a.addr} \smalldeq{= A.addr + i \times (A.width+1) + order}) based on input parameters \textit{order}, \textit{size} and \textit{i}, where \textit{size} contains the matrix dimensions (width and height),  then issues a DMA request to fetch vector \textit{a} from external memory into the SPM (as vector \textit{v}). 
In the \texttt{HOUSE} stage, the shared FP-ALU computes the norm of \textit{v} and combines it with $v[1]$ to generate \textit{q}, which defines the Householder transformation.

\textit{HOUSE\_MM\_UPDATE} is common to both Householder Reduction and Accumulation. 
In the \texttt{VEC DIVISION} stage, the {FP-ALU} calculates the scaling factor \smalldeq{\beta} by multiplying \textit{q} and \smalldeq{v[1]} from the preceding stage (or reading \smalldeq{v[1]} from the SPM, depending on the specific phase).
It then computes each element of \textit{v'} by dividing  the corresponding element of \textit{v} by \smalldeq{\beta} and storing the result back in the SPM.
Finally, the \texttt{REQUEST GEMM} stage issues two consecutive GEMM operations: the first multiplies \textit{v} with a submatrix of \smalldeq{A} (either transposed or not, depending on \textit{order}), while the second multiplies that result by \textit{v'}. 
The accelerator repeats these operations, incrementing \textit{i} and toggling \textit{order} as needed, until the HBD process completes.

\vspace{-2pt}
\subsection{Design of the SORTING and TRUNCATION Modules}
\label{subsec:sort_trunc}

\begin{figure}[t]
    \begin{subfigure}{0.44\linewidth}
        \centering
        \hspace{4mm}
        \includegraphics[width=\textwidth]{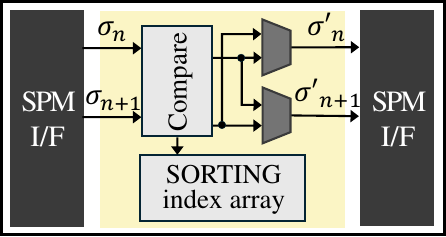}
        \caption{}\label{fig:SORTING}
    \end{subfigure}
    \hfill
    \begin{subfigure}{0.44\linewidth}
        \centering
        \hspace{-4mm}
        \includegraphics[width=\textwidth]{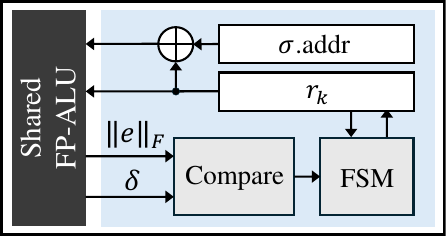}
        \caption{}\label{fig:TRUNCATION}
    \end{subfigure}
    \vskip -4pt
    \caption{Arch. of (a) SORTING and (b) TRUNCATION modules.}\label{fig:SortTrunc}
\end{figure}

In addition to HBD, the TTD-Engine accelerates the \textit{Sorting} and \textit{Truncation} phases of TTD, both of which can incur heavy data movement if executed on the main core. By introducing dedicated modules, we significantly reduce latency from repeated data retrieval.
First, \refFigure{fig:SORTING} shows the {SORTING} module, which implements a bubble-sort-based algorithm for singular values stored in the SPM. 
For each pair of adjacent singular values ($\sigma_{n}, \sigma_{n+1}$), the shared FP-ALU compares them and stores the sorted results ($\sigma'_{n}, \sigma'_{n+1}$) back in the SPM, updating a \textit{SORTING index vector} to track the new order. Once sorting is complete, the module reorders the vectors composing matrices \textit{U} and \textit{V} according to the \textit{SORTING index vector}.

Next, \refFigure{fig:TRUNCATION} illustrates the {TRUNCATION} module, 
which accelerates the $\delta$-\textit{Truncation} procedure and employs a lightweight FSM
to control its operations. At the start of TTD, the module computes the threshold 
$\delta$ by obtaining the Frobenius norm of the input tensor, 
$\delta = \frac{\epsilon}{\sqrt{d-1}}\lVert\bm{W}\rVert_{F}$, 
which is simplified to the norm of the singular values $\sigma$ from the first SVD. 
The shared FP-ALU sequentially performs \textit{SQRT}, \textit{MUL}, and \textit{DIV} 
operations to obtain $\delta$. For each truncation request, the module examines the 
tail of the singular-value vector to form an error vector $\bm{e}$ and checks 
$\|\bm{e}\|_{2} > \delta$. If the condition holds, it updates the truncated 
rank $r_{k}$; otherwise, it decrements $r_{k}$ and repeats the process until the 
desired accuracy is achieved.

\begin{figure}[b]
    \centering
    \includegraphics[width=0.78\columnwidth]{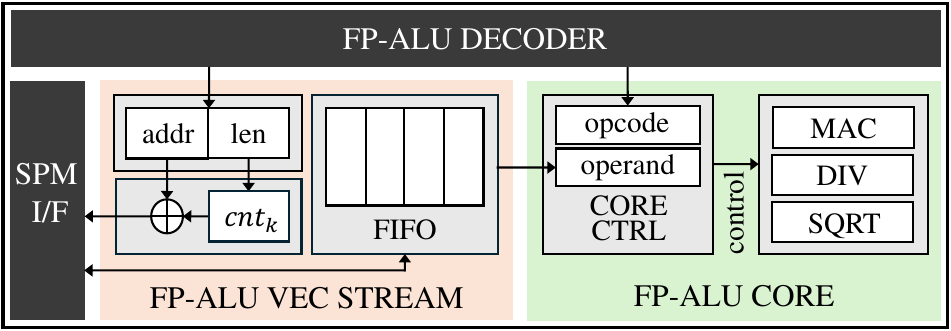}
    \vskip -2pt
    \caption{Architecture of the Shared FP-ALU.}
    \label{fig:FP_ALU}
\end{figure}

\vspace {-4pt}
\subsection{Design of the Shared FP-ALU}
\label{subsec:shared_fp_alu}

Although the GEMM accelerator handles matrix multiplications, TTD still requires a range of additional floating-point operations. 
To reduce latency and hardware overhead, we incorporate a single {Shared FP-ALU} that is shared across the HBD-ACC, SORTING, and TRUNCATION modules. \refFigure{fig:FP_ALU} shows the {Shared FP-ALU}, which comprises an \texttt{FP-ALU Vector Streamer}—responsible for loading and storing vector elements from and to the SPM—and an \texttt{FP-ALU CORE}, containing a customized set of floating-point units ({MAC}, {DIV}, and {SQRT}) derived from an open-source FPU~\cite{Mach:TVLSI20}.

Because TTD frequently computes vector norms, we design the FP-ALU to be able to provide a dedicated \textit{norm} operation. 
When given the opcode and vector address/length, the streamer sequentially reads elements from the SPM into a FIFO, and the \texttt{FP-ALU CORE} squares and accumulates them via \textit{MAC} operations before applying a final \textit{SQRT}. 
It also supports single operations (\textit{ADD}, \textit{MUL}, \textit{MAC}, \textit{DIV}, and \textit{SQRT}) by sending operands directly to the \texttt{FP-ALU CORE}.

By centralizing all floating-point operations in this Shared FP-ALU, we avoid replicating multiple floating-point units across different modules, thereby reducing resource overhead and improving the overall energy efficiency of the TTD-Engine.

\begin{figure}[!t]
    \centering
    \includegraphics[width=1\columnwidth]{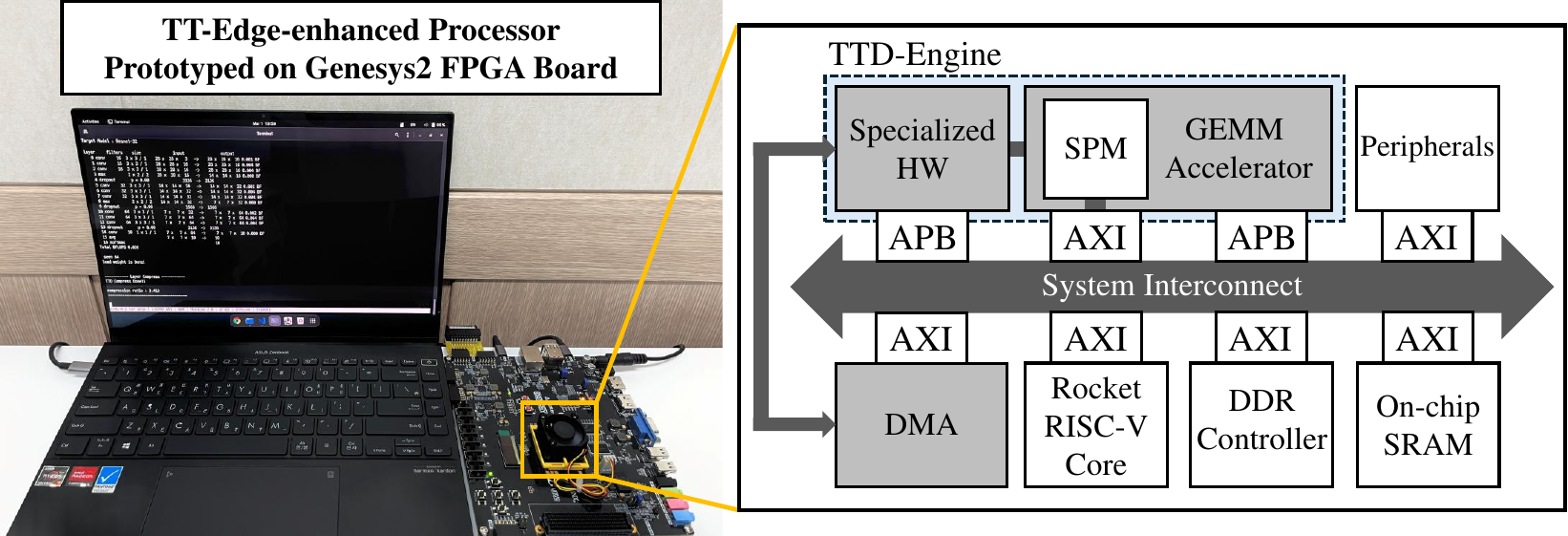}
    \caption{Block diagram of the TT-Edge processor, prototyped on a Genesys2 FPGA board.}
    \label{fig:SoC}
\end{figure}

\begin{table}[b]
    \caption{Resource usage of the TT-Edge processor prototype on Genesys2 FPGA and post-synth. power breakdown at 45 nm.}
    \vskip -2pt
    \centering
    \resizebox{0.98\columnwidth}{!}{
    \renewcommand{\arraystretch}{1.05}
    \begin{tabular}{cll|c|c|c}
    \Xhline{1pt}
    \multicolumn{3}{c|}{{IPs}}            & {LUTs}  & {FFs}       & {Power ($\mathrm{mW}$)\textsuperscript{*}}\\ \Xhline{0.8pt}
    \multicolumn{3}{c|}{Rocket RISC-V Core}      & 15,041         & 9,890    &   10.90 / 2.63\textsuperscript{**}       \\ \hline
    \multicolumn{3}{c|}{SRAM}                    & 166            & 323       &    1.87     \\ \hline
    \multicolumn{3}{c|}{DDR Controller}          & 7,961          & 7,581      &   89.12     \\ \hline
    \multicolumn{3}{c|}{Peripherals incl. DMA}             & 5,047          & 10,373  &    10.60        \\ \hline
    \multicolumn{3}{c|}{System Interconnect}     & 9,748          & 17,376             & 17.78\\ \hline

    \multicolumn{1}{c|}{\multirow{7}{*}{\shortstack{TTD-\\Engine}}}    & \multicolumn{2}{c|}{GEMM Accelerator} & 84,150        & 32,939    & 40.77          \\\hhline{|~-----|}
     \multicolumn{1}{c|}{}     &\multicolumn{2}{c|}{Specialized HW Modules}              & 7,273            & 6,517 &7.19      \\ \cline{2-6}
    \multicolumn{1}{c|}{}      & \cellcolor[HTML]{EFEFEF}$\llcorner$ &\cellcolor[HTML]{EFEFEF}HBD-ACC        & \cellcolor[HTML]{EFEFEF}1,346          & \cellcolor[HTML]{EFEFEF}1,411     & \cellcolor[HTML]{EFEFEF}1.42 \\ 
    \multicolumn{1}{c|}{}     & \cellcolor[HTML]{EFEFEF}$\llcorner$ &\cellcolor[HTML]{EFEFEF}TRUNCATION     & \cellcolor[HTML]{EFEFEF}413            & \cellcolor[HTML]{EFEFEF}884              &\cellcolor[HTML]{EFEFEF}0.78 \\ 
    \multicolumn{1}{c|}{}     & \cellcolor[HTML]{EFEFEF}$\llcorner$ &\cellcolor[HTML]{EFEFEF}SORTING        & \cellcolor[HTML]{EFEFEF}756            & \cellcolor[HTML]{EFEFEF}476                & \cellcolor[HTML]{EFEFEF}0.49\\ 
    \multicolumn{1}{c|}{}     & \cellcolor[HTML]{EFEFEF}$\llcorner$ &\cellcolor[HTML]{EFEFEF}FP-ALU         & \cellcolor[HTML]{EFEFEF}3,314          & \cellcolor[HTML]{EFEFEF}2,287          & \cellcolor[HTML]{EFEFEF}2.23\\
    \multicolumn{1}{c|}{}     & \cellcolor[HTML]{EFEFEF}$\llcorner$ &\cellcolor[HTML]{EFEFEF}Interfaces     & \cellcolor[HTML]{EFEFEF}1,412          & \cellcolor[HTML]{EFEFEF}1,167           & \cellcolor[HTML]{EFEFEF}1.43\\ \Xhline{1pt}
    \multicolumn{6}{r}{\textsuperscript{*}\footnotesize Power analysis obtained from PrimeTime PX.~ \textsuperscript{**}\footnotesize No clock gating / With clock gating.~}
    \end{tabular}
    }
    \label{tab:fpga_and_power}
\end{table}

\section{Experimental Work}\label{sec:exp}
\vspace {-4pt}
\subsection{Implementation}  
We first developed two processor designs: the baseline processor and the TT-Edge processor integrating our proposed TTD-Engine. 
Both designs were implemented at the RTL level in Verilog HDL using RISC-V eXpress (RVX), an EDA tool widely adopted for developing processors on the RISC-V platform~\cite{Han:IoT2021,Park:Date23,Park:TCASI24,Lee:IoTJ25,CHOI:AEJ25}.
The baseline processor consists of a Rocket core supporting floating-point instructions, a DDR3 system memory interface, a 64-PE GEMM accelerator, a DMA engine, and additional peripherals for external interfaces and control. 
In the baseline design, the GEMM accelerator’s control interface employs APB, whereas its data interface is implemented with AXI to ensure low-latency transfers to and from the on-chip scratchpad memory (SPM).
By contrast, the TT-Edge processor replaces the baseline’s GEMM accelerator with our TTD-Engine. 
Its control signals connect to the system interconnect via APB, while dedicated interfaces link the TTD-Engine with the processor’s DMA, SPM, and GEMM hardware, achieving tight integration. \refFigure{fig:SoC} illustrates the overall architecture of the TT-Edge processor.

Next, the TT-Edge processor was prototyped on a Digilent Genesys2 board~\cite{Genesys2} featuring a Xilinx Kintex7-325T FPGA device, operating at 100 MHz. 
\refTable{tab:fpga_and_power} summarizes the resource utilization of each IP block in our processor prototype. 
Focusing first on the specialized hardware modules within the TTD-Engine (excluding the existing GEMM accelerator), we observe that the HBD-ACC—which handles the bulk of TTD computations—consumes 18.5\% of LUTs and 21.7\% of FFs. 
Meanwhile, the SORTING and TRUNCATION modules occupy 10.4\% and 5.7\% of LUTs (and 7.3\% and 13.6\% of FFs), respectively, and the Shared FP-ALU takes up 45.6\% of LUTs and 35.1\% of FFs. 
Furthermore, the DMA, SPM, and GEMM interface logic, along with the associated interconnect, account for 19.4\% of LUTs and 17.9\% of FFs in the TTD-Engine. 
Because all TTD operations share the GEMM accelerator’s existing SPM, the TTD-Engine requires no additional BRAM resources.
From a system-wide perspective, the TTD-Engine introduces minimal overhead by leveraging the existing GEMM accelerator and maintaining a lightweight design for its specialized hardware modules. 
In fact, the additional resources for the TTD-Engine contributes only 5.6\% of LUTs and 7.7\% of FFs across the entire processor, confirming that its resource footprint remains modest.

Finally, we implemented a benchmark application based on ResNet-32 model compression to validate our proposed techniques. 
This application employs TTD with a 3.4$\times$ compression ratio for model parameters. 
We integrated a clock-gating API into the application so that the main core can be clock-gated during HBD, sorting, and truncation—stages handled by the TTD-Engine.
Running this application on our FPGA prototype allowed us to evaluate and validate the effectiveness of the proposed TT-Edge design.

\vspace {-4pt}
\subsection{Verification and Evaluation}

We synthesized the proposed TT-Edge processor using Synopsys Design Compiler under the Nangate 45nm process technology~\cite{NCSU}, and then performed detailed power analysis with Synopsys PrimeTime PX~\cite{PrimeTime}. 
The results are also reported in \refTable{tab:fpga_and_power}. 
Without clock gating on the main core, TT-Edge consumes a total of $178.23\,\mathrm{mW}$, representing only about a 4\% increase relative to the baseline processor’s $171.04\,\mathrm{mW}$. 
This increment is attributable to the specialized hardware modules in the TTD-Engine, excluding the reused GEMM accelerator. 
Breaking down the TTD-Engine’s components, the HBD-ACC contributes $1.42\,\mathrm{mW}$ (19.7\%), while the SORTING and TRUNCATION modules consume $0.49\,\mathrm{mW}$ (6.8\%) and $0.78\,\mathrm{mW}$ (10.8\%), respectively. 
The Shared FP-ALU occupies $2.23\,\mathrm{mW}$ (31\%), and the remaining DMA/\,SPM/\,GEMM interface and interconnect logic draws $1.43\,\mathrm{mW}$ (19.9\%).

\begin{table}
\vskip -5pt
\caption{Execution time \smalldeq{T_{exec}} and energy \smalldeq{E} breakdown for TTD-based ResNet-32 compression on baseline and TT-Edge.}
\vskip -2pt
\centering
\resizebox{0.98\columnwidth}{!}{
\renewcommand{\arraystretch}{1.08}
\begin{tabular}{c|S|S|S|S}
\Xhline{1pt}
\multirow{2}{*}{TTD procedure} &\multicolumn{2}{c|}{Baseline} & \multicolumn{2}{c}{TT-Edge} \\ \cline{2-5}
 & {\smalldeq{T_{exec}} (ms)} 
 & {\smalldeq{E} (mJ)} 
 & {\smalldeq{T_{exec}} (ms)} 
 & {\smalldeq{E} (mJ)} \\ \Xhline{0.8pt}
HBD             & 5626.42 & 962.17  & 2743.80 & 466.34\textsuperscript{*} \\ \cline{1-5} 
QR Decomp.      & 1554.66 & 265.91  & 1554.66 & 277.09 \\ \cline{1-5} 
Sort. \& Trunc. & 312.56  & 53.46   & 31.37   & 5.33\textsuperscript{*}   \\ \cline{1-5} 
Update SVD In.  & 46.65   & 8.15    & 46.65   & 8.49   \\ \cline{1-5} 
Reshape \& etc  & 189.24  & 32.37   & 189.24  & 33.73  \\ \hhline{|-----|}
\cellcolor[HTML]{EFEFEF}Total
& \cellcolor[HTML]{EFEFEF}7729.52
& \cellcolor[HTML]{EFEFEF}1322.06
& \cellcolor[HTML]{EFEFEF}4566.71
& \cellcolor[HTML]{EFEFEF}790.97 \\ \Xhline{1pt}
\multicolumn{5}{r}{\textsuperscript{*}\footnotesize The core is clock gated.}
\end{tabular}
}
\label{tab:ex_time_and_energy}
\end{table}

Notably, the main core can enter a sleep state (i.e., be clock-gated) during the primary TTD compression phases-HBD, Sorting, and Truncation. 
In this scenario, TT-Edge operates at $169.96\,\mathrm{mW}$, which is even less than the baseline’s power consumption. 
This reduction demonstrates the success of our lightweight TTD-Engine design, wherein each submodule (HBD-ACC, SORTING, TRUNCATION) shares a single FP-ALU, and the existing GEMM accelerator is reused to minimize additional hardware requirements.

The performance improvements achieved by TT-Edge are reported in \refTable{tab:ex_time_and_energy}. More specifically, the table compares the execution times of a TTD-based model compression application on both the baseline and TT-Edge processors, divided into HBD, QR decomposition, Sorting \& Truncation, Update SVD Input, and Reshape \& etc. The most time-consuming step, HBD, requires $5626.42\,\mathrm{ms}$ on the baseline—72.8\% of the total TTD runtime—but only $2743.8\,\mathrm{ms}$ on TT-Edge, corresponding to a $2.05\times$ speedup. The Sorting \& Truncation step also shows dramatic gains, dropping from $312.56\,\mathrm{ms}$ on the baseline to $31.37\,\mathrm{ms}$ on TT-Edge, yielding a $9.96\times$ acceleration. As a result, the total TTD runtime is reduced from $7729.52\,\mathrm{ms}$ to $4566.71\,\mathrm{ms}$, representing an overall \textbf{1.7$\bm{\times}$} speedup.  

Furthermore, \refTable{tab:ex_time_and_energy} presents the energy savings achieved by TT-Edge. The table reports the energy consumption of the benchmark application’s main procedures on the baseline and TT-Edge processors, highlighting a 51.5\% reduction during HBD and a 90\% reduction in the Sorting \& Truncation step—both of which clock-gate the main core. Overall, TT-Edge reduces total energy consumption by approximately \textbf{40.2\%} compared to the baseline, demonstrating the effectiveness of our solution in delivering substantial improvements in both performance and energy efficiency.

\begin{table}[t]
\caption{Comparison of the proposed TT-Edge approach with a related technique.}
\centering
\resizebox{0.82\columnwidth}{!}{
\renewcommand{\arraystretch}{1.08}
\begin{tabular}{c|c|c}
\Xhline{1pt}
    Resource Metrics            &{\cite{Qu:TCAD21}}  & {TT-Edge}              \\ \Xhline{0.8pt}
    Process technology          & 45\,{nm}        & 45\,{nm}         \\ \hline
    Number of PEs                & 256 + 64                  & 64 + 3  \\ \hline
    On-chip memory              & 1\,MB                       & 128\,KB + 320\,KB  \\ \hline
    Arithmetic precision        & 16-bit fixed              & 32-bit floating         \\ \hline
    Clock frequency                   & 400\,MHz                    & 100\,MHz \\ \hline
    Power consumption           & 2.89\,{W}            & 48\,{mW} (177\,{mW}\textsuperscript{*})                
    \\ \Xhline{1pt}
     \multicolumn{3}{r}{\textsuperscript{*}\footnotesize Total processor power consumption.}     
\end{tabular}
}
\label{tab:Comp_SOTA}
\end{table}

\section{Discussion}

To date, only limited work has explored TTD-based compression for low-power, resource-constrained edge devices. 
Among early hardware TTD systems, Qu \textit{et al.}~\cite{Qu:TCAD21} designed a dedicated accelerator for compression. 
In contrast, TT-Edge integrates TTD into a complete edge processor through a hardware–software (HW–SW) co-design approach, reusing the existing GEMM accelerator (64 PEs) and adding only three specialized PEs in the TTD-Engine’s shared FP-ALU. 
As a result, the design requires modest on-chip memory (448\,KB total, including a 320\,KB scratchpad inside the GEMM accelerator) and adds just 48\,mW for the TTD-Engine itself (177\,mW for the entire processor), while supporting 32-bit floating-point operations for compatibility with PyTorch~\cite{Pytorch} and TensorFlow~\cite{TensorFlow}. 
Taken together, \refTable{tab:Comp_SOTA} highlights that TT-Edge is well suited for low-power edge scenarios demanding both efficiency and floating-point capability.

While TT-Edge effectively alleviates the major latency bottlenecks of TTD on edge AI,
some data movement overhead between DRAM and the SPM remains.
Our design mitigates this cost (e.g., by storing Householder vectors on-chip), and further reductions could be achieved by exploring alternative memory hierarchies and interconnects.
We view such directions—including advanced interconnects, memory subsystems, or near-memory processing—as promising opportunities for future work to further enhance TT-Edge’s efficiency.

\section{Conclusion}

We presented \emph{TT-Edge}, a HW–SW co-designed framework that accelerates TTD-based model compression on resource-limited edge AI processors. 
By splitting SVD into Householder-based bidiagonalization and QR-based diagonalization, and offloading the heavy computations to a lightweight dedicated accelerator, TT-Edge achieves efficient compression with modest hardware cost. 
The TTD-Engine is tightly integrated with the existing GEMM unit, enabling resource reuse and reducing matrix–vector transfers, which substantially lowers both latency and energy consumption. 
Implemented on a RISC-V-based processor prototype, TT-Edge achieves a 1.7$\times$ speedup and a 40.2\% reduction in energy consumption compared to the baseline, with only a 4\% power overhead. 
These results demonstrate that advanced tensor decompositions can be made practical even under the constraints of low-power edge devices.



\bibliographystyle{IEEEtran}
\bibliography{reference}
\end{document}